# Exchange-Mediated Mutual Correlations and Dephasing in Free-Electrons and Light Interactions

Nahid Talebi[1,*] and Iva Březinová[2]

[1]Institute for Experimental and Applied Physics, Christian Albrechts University, Leibnizstr. 19, 24118 Kiel, Germany

[2]Institute for Theoretical Physics, Vienna University of Technology, Wiedner Hauptstrasse 8-10/136, Vienna, Austria

E-mail: talebi@physik.uni-kiel.de

**Abstract** – The quantum world distinguishes itself from the classical world by being governed by probability amplitudes rather than probabilities. On a single-particle level, quantum phases can be manipulated leading to observable interference patterns that can be used as a probe e.g. in matter wave microscopy. But the quantum world bears even more fascinating effects when it comes to the interplay between more than one particle. Correlations between quantum particles such as entanglement can be exploited to speed up computational algorithms or enable secure cryptography. Here, we propose and numerically explore a thought experiment to address the question whether quantum correlations between particles can be used in matter wave microscopy. Specifically, we address the following questions: Can information be transferred between two mutually spin-correlated free-electron wavepackets? Can Coulomb and exchange correlations be linked to the decoherence and dephasing mechanisms of matter waves? Using a time-dependent Hartree-Fock algorithm, we will show that the exchange term has a substantial role in transferring the information between two mutually spin-correlated electrons, whereas the Hartree potential (or mean-field Coulomb potential) dominates the dephasing on a single-particle level. Our findings might facilitate fermionic matter-wave interferometry experiments designed to retrieve information about non-classical correlations and the mechanism of decoherence in open quantum systems.

**Introduction**

In contrast to classical probability distributions, quantum probabilities are determined by probability amplitudes. The ability to coherently manipulate the phase of a quantum object with holograms or laser light, and also to detect it, has revolutionized the world of matter-wave interferometry [1, 2] and microscopy [3]. After pioneering experiments by Boersch demonstrating the diffraction of coherent electron beams by macroscopic objects [4], holography was proposed by Gabor in 1948 [5], as a tool for improving the spatial resolution in electron microscopy. Nowadays, technological advancements in aberration corrected electron microscopy [6, 7] have enabled a resolution far beyond what could be conceived in 1948.

The ability to coherently manipulate the phase of a free-electron wavepacket using near-field optical distributions in the vicinity of nano-objects has been manifested by Zewail and coworkers [8], pioneering the field of photon-induced near-field electron microscopy (PINEM) [9]. Ground-breaking experiments carried out by Ropers and coworkers have demonstrated that Rabi oscillations in the energy-momentum ladders induced by the laser field lead to the formation of attosecond electron bunches [10]. Moreover, strong laser-photon interactions [11-13] might ultimately lead to entangled electron-photon states [14]. In addition, PINEM can be used to holographicaly recover optical wave fronts [15], and for quantum state tomography of optical states [16] in an inverse approach. In addition to PINEM, coherent manipulation of the electron phase by transverse light in free space due to nonlinear processes caused by the ponderomotive interaction [17, 18] paves the way for on-demand electron-wave shaping and might be used for phase-contrast microscopy.

Theoretical understanding of electron-light interactions, particularly within the context of PINEM and free-space processes has significantly benefited from eikonal and adiabatic approximations, providing a fully analytical platform for interpreting experimental results [9, 19]. The adiabatic approximation is a sound basis for high-energy electrons interacting with low-energy optical waves. However, a significant domain of physical processes is not covered by this approximation, as it neglects amplitude modulations since recoil and diffraction processes cannot be modelled [20]. This domain – typically called the non-adiabatic domain – can be addressed numerically using a Maxwell-Schrödinger numerical toolbox [21]. Particularly, it has been shown that Kapitza-Dirac diffraction, occuring when the optical excitations are phase-matched with the electron wavepacket motion, can be used as a probe of quantum coherence in diffraction experiments [12]. The visibility of such diffraction patterns unambiguously determines the mutual coherence between the field and the electron wavepackets.

In the investigations stated above, the fermionic statistics of the electrons did not play any role. As to what extent the electrons behave differently compared to bosons like photons in matter-wave interferometry and PINEM experiments has not yet been addressed to the best of our knowledge. Spin-polarized electron waves obtained from GaAs field-emission sources [22, 23], in principle, provide a platform to investigate the effects of the fermionic statistics. The outcomes of matter-wave experiments taking into account the fermionic statistics and the Pauli exclusion principle (PEP) are conceptualized in this report using numerical simulations. The simplest theory correctly accounting for the PEP in many-electron systems is the Hartree-Fock (HF) theory. Here, a thought experiment is devised and its outcomes are numerically explored by extending the afore-mentioned Maxwell-Schrödinger toolbox by the time-dependent HF (TDHF) method [24]. We consider a simple system including two electron wavepackets both with parallel and anti-parallel spin polarizations interacting with the laser-induced near-field optical distribution of a gold nanorod. We investigate their quantum coherent optical phase modulations and their mutual interactions mediated by both Coulomb mean-field and exchange terms. It will be shown that the exchange potential facilitates an exchange of phase information between the electron wavepackets. Our findings, pave the way towards matter-wave experiments beyond the routinely employed unpolarised electron systems, i.e. experiments where fermionic statistics is exploited to retrieve and investigate the transfer of information between entangled electron wavepackets.

**Results and Discussion**

**Time-dependent Hartree-Fock Theory and Exchange Correlations**

To set the stage, we consider first an *N*-electron system interacting with a laser field. Within TDHF the ansatz for the N-electron wavefunction corresponds to a Slater-determinant consisting of N different single-particle states (or HF orbitals). The equations of motion for each orbital within the minimum-coupling Hamiltonian can be written as

$$\left[\frac{1}{2m_0}\left(\hat{\vec{p}}+e\vec{A}(\vec{r},t)\right)^2 + \hat{v}^H + \hat{v}_k^x - e\varphi(\vec{r},t)\right]\psi_{ns_n}(\vec{r},t) = i\hbar\frac{\partial}{\partial t}\psi_{ns_n}(\vec{r},t) \quad , \tag{1}$$

where $\hat{\vec{p}} = -i\hbar\vec{\nabla}$ is the momentum operator, $\vec{A}(\vec{r},t)$ and $\varphi(\vec{r},t)$ are the time-dependent vector and scalar potentials, respectively, $m_0$ is the electron mass, and $\hbar$ is the reduced Planck constant. $\psi_{ns_n}$ are the individual single-electron states (here propagating wavepackets). The subscripts $n$ and $s_n$ denote the electron index and the spin degree of freedom of the $n^{th}$ orbital, respectively. $\hat{v}^H$ and $\hat{v}_k^x$ are the Hartree and exchange operators, expressed as

$$\hat{v}^H \psi_{ns_n}(\vec{r},t) = \frac{e^2}{4\pi\varepsilon_0}\sum_{j\neq n}\int d^3r' \frac{\psi_{js_j}(\vec{r}',t)\psi^*_{js_j}(\vec{r}',t)}{|\vec{r}-\vec{r}'|}\psi_{ns_n}(\vec{r},t) \tag{2}$$

and

$$\hat{v}_n^x \psi_{ns_n}(\vec{r},t) = -\frac{e^2}{4\pi\varepsilon_0}\sum_{m\neq n}\delta_{s_m s_n}\int d^3r' \frac{\psi^*_{ms_m}(\vec{r}',t)\psi_{ns_n}(\vec{r}',t)}{|\vec{r}-\vec{r}'|}\psi_{ms_m}(\vec{r},t) \quad , \tag{3}$$

respectively. $\varepsilon_0$ is the free-space permittivity, and $\delta$ is the Kronecker-delta function. For quasi-free electron wavepackets with peak kinetic energy of $\hbar\omega_n = \hbar^2 k_n^2/2m_0$, it is beneficial to recast the wave function as $\psi_{ns_n}(\vec{r},t) = \tilde{\psi}_{ns_n}(\vec{r},t)e^{i\vec{k}_n\vec{r}-i\omega_n t}$, and expand eq. (1) as

$$\frac{-\hbar^2}{2m_0}\left(\nabla^2\tilde{\psi}_{ns_n} + i\vec{k}_n\cdot\vec{\nabla}\tilde{\psi}_{ns_n}\right) - \frac{i\hbar e}{m_0}\vec{A}\cdot\left(\vec{\nabla}\tilde{\psi}_{ns_n} + i\vec{k}_n\tilde{\psi}_{ns_n}\right)$$
$$+\left(\hat{v}^H - e\varphi(\vec{r},t)\right)\tilde{\psi}_{ns_n}(\vec{r},t) + \hat{v}_n^x\left[\tilde{\psi}_{ns_n}(\vec{r},t)\right] \tag{4}$$
$$= +i\hbar\frac{\partial}{\partial t}\tilde{\psi}_{ns_n}$$

We have used Coulomb gauge in the above equation. For a system including spin-polarized electrons with a prescribed spin orientation, eq. (4) can be explicitly written as a coupled system of equations given by

$$\frac{-\hbar^2}{2m_0}\left(\nabla^2\tilde{\psi}_n + i\vec{k}_n\cdot\vec{\nabla}\tilde{\psi}_n\right) - \frac{i\hbar e}{m_0}\vec{A}\cdot\left(\vec{\nabla}\tilde{\psi}_n + i\vec{k}_n\tilde{\psi}_n\right)$$
$$+\left(\sum_{m\neq n}\tilde{v}_m^H - e\varphi(\vec{r},t)\right)\tilde{\psi}_n(\vec{r},t) + \sum_{m\neq n}\tilde{v}_{nm}^x\tilde{\psi}_m(\vec{r},t) \quad , \tag{5}$$
$$= +i\hbar\frac{\partial}{\partial t}\tilde{\psi}_n$$

with $n \neq m$. The Hartree and exchange potentials are recast as

$$\tilde{v}_m^H = \frac{e^2}{4\pi\varepsilon_0} \int d^3r' \frac{|\tilde{\psi}_m(\vec{r}',t)|^2}{|\vec{r}-\vec{r}'|} \tag{6}$$

and

$$\tilde{v}_{nm}^x = -\frac{e^2}{4\pi\varepsilon_0} e^{i(\vec{k}_m-\vec{k}_n)\cdot\vec{r}'} \int d^3r' \frac{\tilde{\psi}_m^*(\vec{r}',t)\tilde{\psi}_n(\vec{r}',t) e^{-i(\vec{k}_m-\vec{k}_n)\cdot\vec{r}'}}{|\vec{r}-\vec{r}'|}. \tag{7}$$

Reformulating equation (1) as (5), allows to limit the size of the simulation domain in the momentum space and facilitates the use of the Fourier method for spatial derivations [25]. Clearly, the exchange potential depends on the phase differences between individual electron wavepackets in the system. In order to conceptualize the exchange-mediated phase correlations, we analyse our two-electron system (Fig. 1) first within the weak and then within the strong-coupling regime between the two electron wavepackets. Using equations (6) and (7), and neglecting for this analysis the kinetic term in order to extract the role of the Coulomb mean-field and exchange term, we derive the following equation of motion for the second orbital (see the Methods section for details) in the limit of weak coupling between the electrons:

$$\hbar \frac{\partial}{\partial t}|\tilde{\psi}_2|^2 = 2\,\text{Im}\frac{e^2}{4\pi\varepsilon_0} C(\vec{r},t) \int d^3r' \frac{C(\vec{r},t)}{|\vec{r}-\vec{r}'|} \tag{8}$$

with

$$C(\vec{r},t) = \tilde{\psi}_1^*(\vec{r},t)\tilde{\psi}_2(\vec{r},t), \tag{9}$$

defined as the time-dependent mutual correlation function between the electron wavepackets. The electron wavepacket will thus experience time-dependent intensity modulations which depend on $C(\vec{r},t)$. In other words, phase differences between the electron wavepackets result in a dynamic change of the absolute square of the probability amplitude, i.e. the particle density. Note that eq. (8) is a nonlinear eigenvalue problem that needs to be solved self-consistently. In case of strong interactions, where $\tilde{\psi}_1$ and $\tilde{\psi}_2$ substantially overlap in space, the dynamic exchange of phase information between the wavepackets significantly modulates the energy-momentum distribution of both wavepackets. The spatial overlap of the wavepackets as a function of time is the parameter that quantifies the strength of the interaction. Since laser-induced near-field distributions strongly modify the phase of the nearby electron wavepacket, mutual phase correlations between the electron wavepackets in a devised experiment as shown in Fig. 1 can be examined by virtue of controlling the distance between the wavepackets ($d_1$) and between the wavepackets and the sample ($d_2$). Nevertheless, the analytical treatment of this system, even including only two wavepackets, is challenging. Therefore, hereafter only numerical simulations will be presented.

**TDHF simulations of spin-polarized and spin-unpolarized electron wavepackets**

We have modified a recently developed numerical toolbox based on the combined Maxwell-Schrödinger equations to include the TDHF formalism and to simulate the interaction of free-electron wavepackets with light within the semi-classical approach [12, 21, 26]. The spatial symmetries of the system allow to restrict oneself to two dimensions (2D; $\vec{r} = (x, y)$). Details can be found in the Methods section. Two initially Gaussian electron wavepackets at the full-width at half-maximum (FWHM) longitudinal and transversal broadenings of 33.2 nm and 3.3 nm, respectively, and at kinetic energies of 1436 eV ($v_1 = 0.0748c$) and 1424 eV ($v_2 = 0.0745c$), respectively, propagate along the x-axis through the interaction medium. The electron impact parameters are taken to be $d_1 = 5\,\text{nm}$ and $d_2 = 20\,\text{nm}$. The interaction medium is composed of a gold nanorod with the radius of 15 nm excited by a pulsed laser field. The *x*-polarized laser pulse has a center wavelength of 800 nm, a FWHM temporal broadening of 30 fs, and its field amplitude is $E_0 = 5\times10^9\,\text{Vm}^{-1}$. The electron wavepacket which propagates at the distance of 5 nm away from the surface of the nanorod interacts strongly with the excited plasmons. The second wavepacket propagates at the distance of 20 nm away from the surface and experiences a weak coupling strength to the excited evanescent tail of the plasmons. This is due to the strongly confined mode volume of the dipolar plasmon excitations. We assume now that both electrons have the same spin – referred to as spin-polarized beams, and their dynamics is thus mediated by both the Coulomb mean-field and exchange potential, as well as the plasmon-mediated electron-photon interactions. The complete state of the two-electron system in Hartree-Fock approximation is given by

$$|\psi\rangle = \frac{1}{\sqrt{2}}(|12\rangle - |21\rangle)|\uparrow\uparrow\rangle \tag{10}$$

where we have used the Dirac notation for spatial orbital one, $\psi_1(\vec{r}, t) \to |1\rangle$, and spatial orbital two, $\psi_2(\vec{r}, t) \to |2\rangle$. Equation (10) satisfies the PEP. Assuming that upon measurement the spin of the electrons is not measured their state is described by the following reduced density matrix ($\rho_{12} = \text{Tr}_s\{|\psi\rangle\langle\psi|\}$)

$$\rho_{12} = \frac{1}{2}(|12\rangle\langle 12| + |21\rangle\langle 21| - |12\rangle\langle 21| - |21\rangle\langle 12|) \tag{11}$$

which still corresponds to a pure two-electron state. This will be different in the spin-unpolarized case where tracing out the spin degrees of freedom leads to a mixed state already on the two-particle level. The last two terms can be associated with the non-vanishing exchange as they are not present in case of unpolarized electrons, see below. To quantify the exchange contribution to the two-electron density matrix in both coordinate and momentum space we evaluate in addition

$$\rho_x = -\frac{1}{2}(|21\rangle\langle 12| + |12\rangle\langle 21|). \tag{12}$$

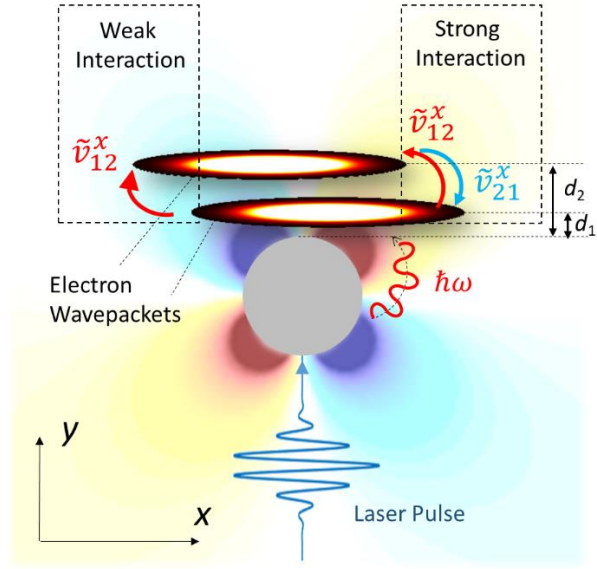

**Figure 1.** Schematic picture of the investigated system consisting of two electron wavepackets interacting with laser-induced plasmonic oscillations. Electrons can interact either strongly or weakly depending on the distance between the electrons $d_2 - d_1$ and the coupling strength mediated by the exchange and Coulomb mean-field interactions. Exchange correlations lead to quantum phase transfer between the wave packets. In the case of weak interactions, the phase of the electron that is nearer to the nanostructure is strongly modified. This phase will be transferred to the second electron by the exchange interaction. Strong-coupling caused by a significant spatial overlap of the wave functions results in a coherent energy and phase exchange between the electrons and significantly alters both wavepackets.

Note that different definitions exist in literature as to the identification of the exchange contribution in two-electron density matrices, see e.g. [27] and the Methods section. In order to underline the differences between the spin-polarized and spin-inpolarized case, we have restricted $\rho_x$ to the part which is present in the first and absent in the latter case.

Selected snapshots of the individual electron wavepackets and the spin-reduced density matrix of the whole two-electron system for several interaction times are shown in Figure 2. The single-particle density matrix is obtained by further tracing out one electron and is given by

$$\rho_1 = \frac{1}{2}(|1\rangle\langle 1| + |2\rangle\langle 2|). \tag{13}$$

nd the particle density is given by the sum of the absolute squares of the two wavepackets. The electronic state at the single-particle level is thus an incoherent mixture of the two orbitals. The source of this decoherence is the interaction between the two electrons. However, each of the orbitals is coherent such that diffraction experiments would lead to an incoherent sum of two high visibility diffraction patterns. This visibility might, however, further be reduced if additional decoherence sources are present such as in [28-31].

Both amplitude and phase of the electron wavepackets are modulated by virtue of their interactions with the near-field distributions. The coupling strength between the laser-field and a single-electron wavepacket in 2D, namely the so-called $g$-factor, is specified by $g = (e/2\hbar\omega_{\text{ph}})\int dk_y \, \tilde{E}_x(k_x = \omega_{\text{ph}}/v_i, k_y; \omega_{\text{ph}})$ [12], where $\omega_{\text{ph}}$ is the photon angular frequency and $v_i$ is the electron velocity. Electrons can inelastically interact with the electric field projected along their trajectory with the energy-momentum conservation being formulated in the form of a selection rule as $k_x = \omega_{\text{ph}}/v_i$. For the wavepacket $\psi_1$ closer to the nanorod at the kinetic energy of 1436 eV, $k_x$ is equal to $k_x = 3.05 k_0$, where $k_0$ is the free-space wavenumber of the photons. Using localized plasmonic modes, this selection rule can be perfectly satisfied at the vicinity of the structure. The strong interaction of the first wavepacket with the plasmonic near-field distribution significantly alters its longitudinal and transverse momentum. The second wavepacket at the distance of 20 nm from the nanorod, only weakly interacts with the near-field light and its PINEM spectrum occupies only a few photon energies (up to $\pm 5\,\hbar\omega$). Nevertheless, the exchange term transfers additional phase information from the first electron to the second causing additional modulation of its PINEM spectrum. Within the interaction time of approximately 5 fs, the two-electron system has reached its largest momentum span (Fig. 2j, k, l, at $t$ = 9.25 fs). Importantly, the classical electron recoil caused by the Lorentz force determines the span of the electron wavepacket in the momentum representation, whereas quantum-mechanical phase modulations, occuring at the modulus of $\hbar k_x = \hbar\omega_{\mathrm{ph}}/v_i$, appear as a longitudinal energy comb, where the energy distance between the ladders is given by the photon energy of $\hbar^2 k_x^2/2m_0 = \hbar\omega_{\mathrm{ph}}$ [12]. Exactly such optical phase modulations lead to the appearance of off-diagonal peaks in the density matrix and significantly modify the spatial and momentum distributions of the two-electron state (compare Fig. 2c and f to Fig. 2 o and r). At longer times we observe a grid like pattern to emerge in the spatial distribution where the probability to find electron pairs with $x_1$ and $x_2$ close to each other is strongly suppressed due to exchange, compare Fig. 3c and Fig. 3g. The reflection symmetry of the density matrix signifies the entanglement between energetically distinct components of the electron wavepackets. The diagonal terms are exactly zero, as expected from the PEP for the spin-polarized two-electron system. The exchange density matrix as well reveals that the probability of having similar momenta is substantially suppressed for the two electrons. This means that the phases of the electrons become increasingly asynchronous over time leading to dephasing observed in the PINEM spectra, see below.

The stripe-like phase modulations cannot be modeled by classical means [12, 29]. In contrast, the span of the energy spectrum (Fig. 3) and the longitudinal momentum are both results of classical interactions. Strong electron-light interactions caused by a larger g-factor also cause distinguished diffraction peaks along the transverse direction that cannot be observed in the weak-interaction regime. At a given time after the interaction ($t = 20$ fs; 10.5 fs after the center of the wavepacket has reached the center of the rod), the wavepackets have completely left the near-field interaction medium such that their energy-momentum distributions are not altered by the electron-photon interactions anymore. The PINEM spectrum, or the so-called electron energy-gain spectrum, can be calculated from the expectation value of the kinetic operator. The angle-resolved differential energy expectation value is represented by [12]

$$\sigma_i(E,\varphi) = \frac{d}{dEd\varphi}\langle\psi_i(x,y,t)|\hat{H}_K|\psi_i(x,y,t)\rangle = \frac{m_0}{\hbar^2}E|\tilde{\psi}_i(E,\varphi,t)|^2, \tag{14}$$

where $i = 1, 2$, $\varphi = \tan^{-1}(k_y/k_x)$, $E = \hbar^2(k_x^2 + k_y^2)/2m_0$, and $\hat{H}_K = \hat{p}^2/2m_0$ is the kinetic energy operator. Thus, the PINEM spectrum for a single-electron wavepacket is calculated as $\Sigma_i(E) = \int_{-\varphi_0}^{+\varphi_0} d\varphi\,\sigma_i(E,\varphi)$, where the span of the angular integration is given by the spectrometer acceptance angle. Here, we integrate over the complete angular span of the simulations, i.e. $\pm 10°$. For the two-electron system discussed here the PINEM spectrum is given by $\Sigma(E) = \Sigma_1(E) + \Sigma_2(E)$ as it is determined from the expectation value of a sum over single-particle operators, i.e. the kinetic energy, and thus is determined from the single-particle density matrix $\rho_1$ (see eq. (13)). The PINEM spectrum as shown in Fig. 3i, features a series of energy combs up to $\pm 60\,n\hbar\omega$, an energy split of the order of $\delta E = 34\,\hbar\omega$ (blue horizontal arrow in Fig. 3 i), and a modulated envelope (vertical blue arrows in Fig. 3i). The fringes have a markedly reduced visibility as compared to the un-polarized electron system discussed below (see Fig. 4 g). The PINEM spectrum is also asymmetric in the energy loss and gain sides. The latter phenomenon is attributed to the diffraction the electron beam experiences, as this cannot be observed for a one-dimensional electron model.

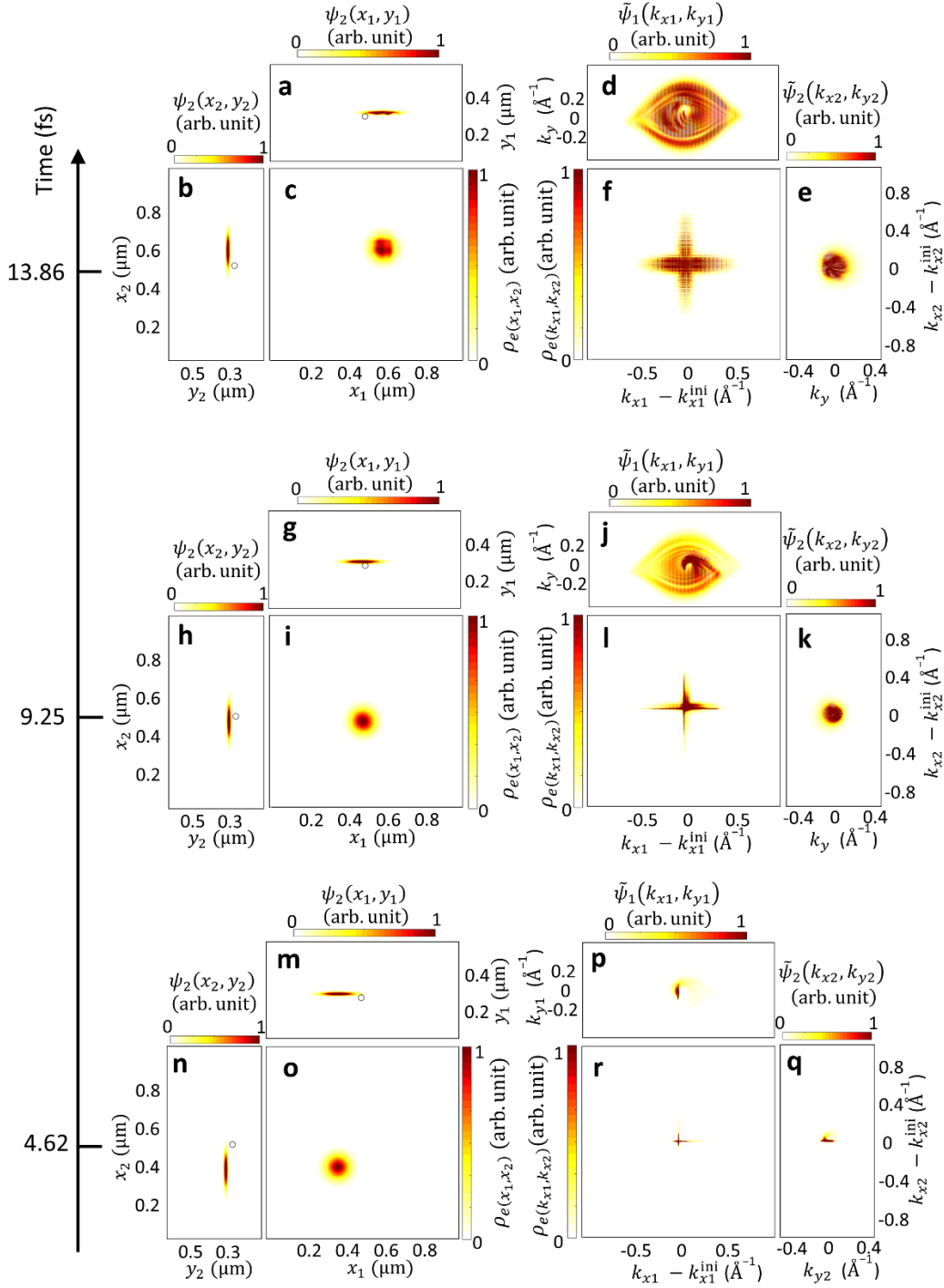

**Figure 2.** Dynamics of the interaction of a spin-polarized two-electron system with the plasmonic excitations of a gold nanorod induced by a coherent and classical light (see the text for a detailed description of the system). Demonstrated are the spatial (c, i, o) and momentum-space (f, l, r) distributions of the spin-reduced density matrices integrated over $y$ or $k_y$, respectively, (see text) at depicted times. In addition, individual wavepackets (HF orbitals) in real (a, b, g, h, m, n) and momentum (d, e, j, k, p, q) space for the first (a, d, g, j, m, p) and second (b, e, h, k, n, q) electron, initially propagating at the distances of 5 nm and 20 nm away from the surface of the rod, respectively, are also depicted. The rim of the nanorod is depicted with a circle in panels a, b, g, h, n, and m.

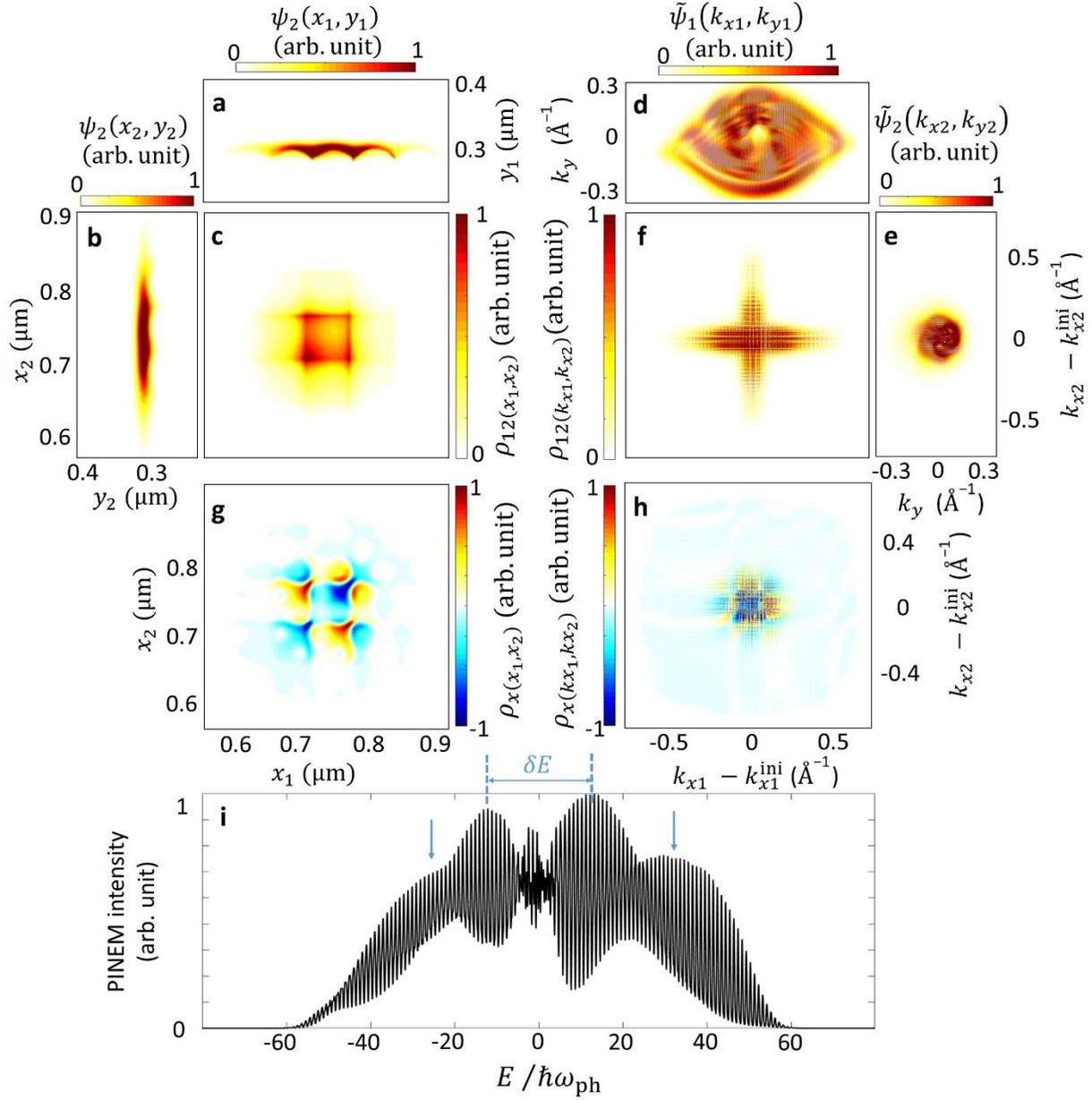

**Figure 3.** Final distributions of the reduced density matrices of a polarized two-electron system interacting with the plasmons induced by a classical and coherent light excitation, and after tracing out spin and the $y(k_y)$-axes (for parameters, see text), within real space (c, g) and momentum space (f, h) at $t = 18\,fs$. (a) and (b) show the spatial distributions of individual wavepackets (orbitals) and (d) and (e) show them in momentum space. Exchange contributions to the density matrices in real (g) and momentum (h) space. (i) Calculated PINEM spectrum.

The overall shape of the probability distributions of the wavepackets in case of an unpolarized two-electron state is similar to the case of polarized wavepackets (compare Fig. 4 with Fig. 3). The overall span of the wavepackets in both the momentum and the real-space domains is practically the same. However, the density matrix distribution especially in momentum space shows marked differences. The density matrix for the unpolarized two-electrons system, after tracing out the spin degrees of freedom, is given by an incoherent sum over two product states, i.e. $\rho_{12} = \frac{1}{2}(|12\rangle\langle 12| + |21\rangle\langle 21|)$, in other words it is a mixed state. The dynamics of this system is not affected by exchange correlations. For completeness we mention that the single-particle state $\rho_1$ is given by the same formula (Eq. 13) as in the polarized case.

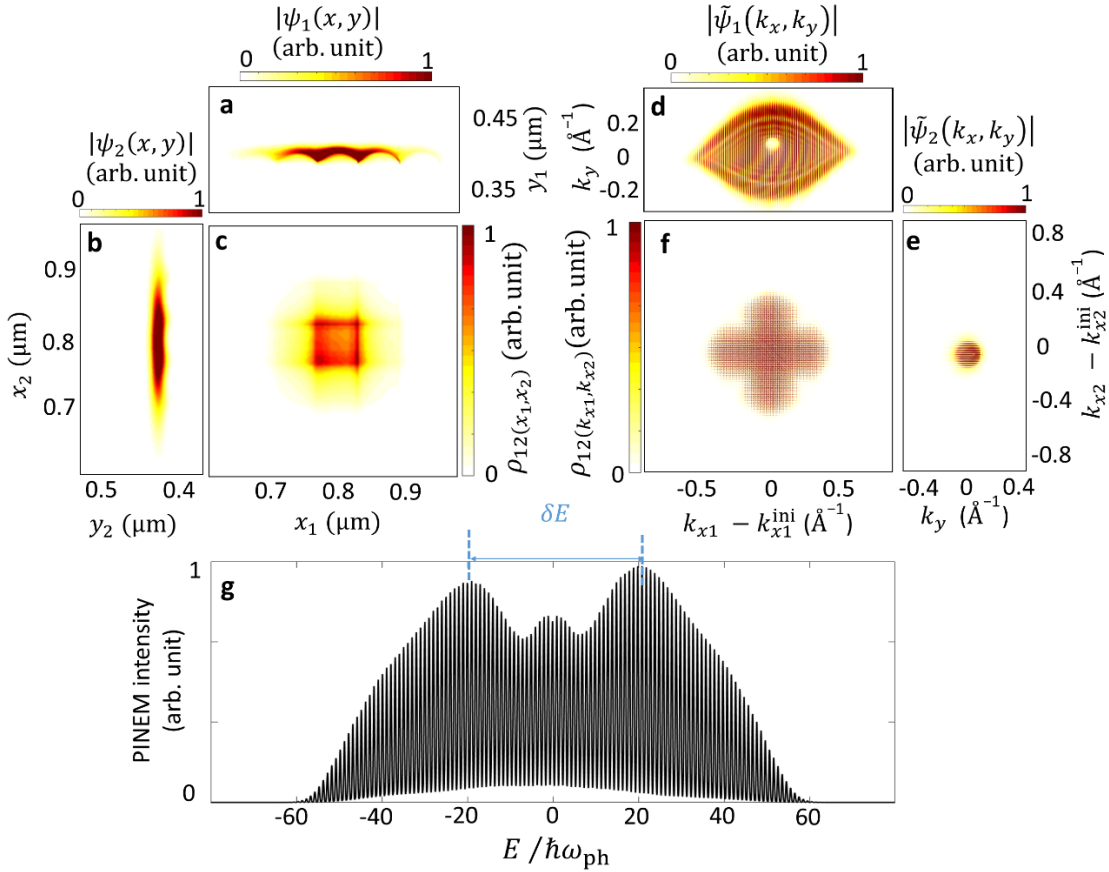

**Figure 4.** Final distributions of the reduced density matrices of an unpolarized two-electron system after interaction with the plasmons induced by a classical and coherent light excitation, and after tracing out the $y(k_y)$-axis, (for parameters, see the text), within the real space (c) and momentum space (f). (a) and (b) show the spatial distributions of individual wavepackets (orbitals) and (d) and (e) show the same but in momentum space. (g) Calculated PINEM spectrum.

Since exchange correlations are not present in this case, the electrons interact only via the Coulomb mean-field potential and no phase information is transferred. The Coulomb interactions modeled by the Hartree potential do not cause any change in the visibility of the PINEM fringes. The PINEM spectrum shows the same high visibility pattern as if only one electron were present. The continuum of momenta (or energy) modulations between individual wavepackets that develops in the polarized case due to exchange of phase information is not present here. The fact that the visibility of the fringes in PINEM spectrum is not much affected by the Coulomb potential is mainly due to the large distance of 15 nm between the electrons.

In the following, we turn back to the polarized case and analyse in more detail the dynamic exchange of phase information between the wavepackets in this case. We focus now on times after the interaction with the laser field has taken place. We consider the same system as described above with the only difference being that the impact parameter for the second electron is now reduced to $d_2 = 10\,\text{nm}$. After the phase information has been imprinted on the electron wavepackets by means of the laser interactions, a dynamic mutual interaction mediated by the exchange correlations results in a continuous modulation of the PINEM spectra associated with each electron wavepacket (see Fig. 5 a and b). The stronger phase exchange present here as compared to the previous example is due to two concurrent effects: (i) the smaller distance of the second electron wavepacket to the nanorod causes a stronger electron-light coupling that leads to the occupation of higher energy gain and loss channels (compare with

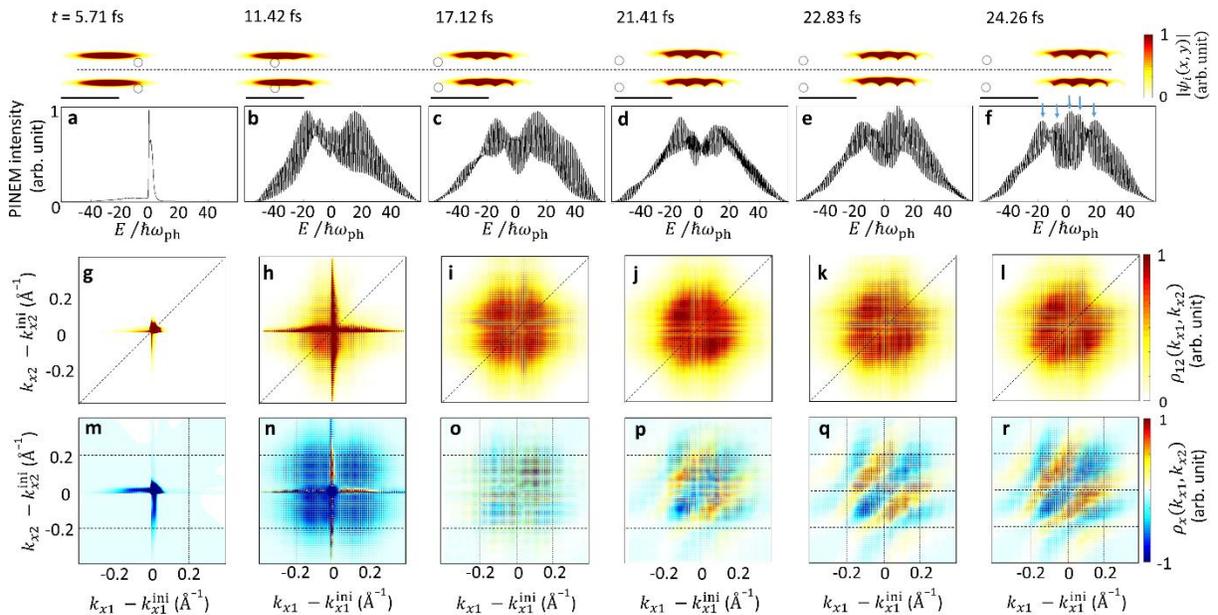

**Figure 5.** Dynamics of the interaction of a spin-polarized two-electron system with the plasmonic excitations of a gold nanorod induced by a coherent and classical light (see the text for a detailed description of the system). The first and second wavepackets traverse the near-field zone at the distances of 5 nm and 10 nm from the surface of the nanorod, respectively. (a-f) PINEM spectra calculated at depicted interaction times above each panel. The insets demonstrate the spatial distributions of individual packets. (g-l) and (m-r) show the total and exchange components of density matrices at each time, respectively.

Fig. 3e); (ii) the larger spatial overlap of the wave packets causes more significant coherent energy-transfer dynamics which leads to distinguished peaks in PINEM (see blue arrows in Fig. 5F) that are less obvious in the previous polarized system and not present in the unpolarised case. The mechanisms of energy transfer is taking place at an ultrafast femtosecond time scale. The exchange correlations thus lead to coherent energy transfer dynamics between the wavepackets, the rate of which depends on the spatial overlap of the wave functions. Moreover, the competing photon and exchange mediated interference paths lead to less visible PINEM fringes – the phenomenon that underlies the dephasing mechanism.

The exchange-mediated correlation effects can be further controlled by tuning the initial center kinetic energies of the individual electron wavepackets. More precisely $\delta \vec{k} = \vec{k}_m - \vec{k}_n$ has a pronounced effect on the energy exchange between the wavepackets. Furthermore, tuning the center energy has the advantage that an ultrafast deflector, like a THz streak camera [30], can be used to individually detect each wavepacket using electron spectrometers (Fig. 6c). An experimental setup for precisely aligning the wavepackets for achieving space-time overlap of the wave functions at the interaction zone and individually detecting each wavepacket could become feasible through photoemission from nanotips, allowing for realization of ultrashort wavepackets. A combination of the photoemission electron sources with magnetic field deflectors and optical delay lines could be used to align the electron wavepackets to achieve the required spatio-temporal overlaps at the interaction site (Fig. 6a and b). The individual detection of the electrons could facilitate a correlated detection and the observation of anti-correlations [32] due to the PEP.

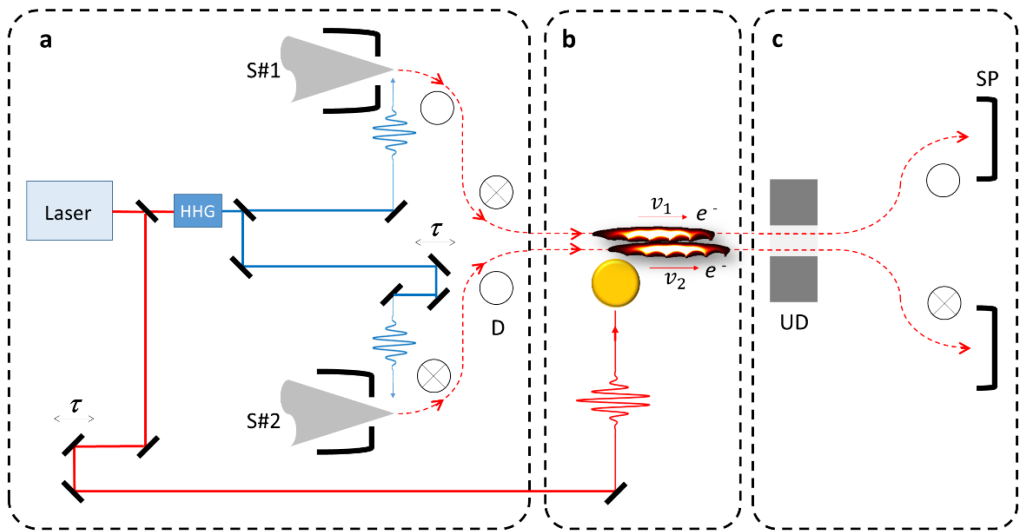

**Figure 6.** An experimental ultrafast point-projection electron microscopy setup allowing to realize spin-polarized electron sources (S#1 and S#2) for investigating the exchange correlations. (a) Two electron wavepackets are accelerated after the photoemission process to reach the centre velocities of v1 and v2, and are aligned using magnetic-field deflectors (D). The delay between the wavepackets is tuned to allow for spatiotemporal overlap of the wavepackets at the interaction zone (b). (c) An ultrafast deflector (UD) deflects the wavepackets into different trajectories, allowing for individually detecting them using a spectrometer (SP).

## Conclusion

In summary, on the basis of realistic numerical time-dependent Hatree-Fock simulations we have investigated the role of exchange correlations in mutual interactions between spin-polarized and spin-unpolarized electron wavepackets. We have shown that the exchange potential causes a coherent transfer of phase information and energy between the wavepackets. The prerequisite is a significant overlap of the wavefunctions and is controlled further by the energy difference between the wavepackets. The initial phase information that is imprinted on the wavefunctions by ultrafast coherent electron-photon interactions is transferred mutually between the wavepackets and therefore can be used to investigate the exchange-mediated coherent energy transfer. Due to the mutual exchange interactions visibility of PINEM fringes of individual wavepackets decreases – a phenomenon that can be linked to a dephasing mechanism. It gives rise to a broad continuum of energies for each wavepacket which might be used in future to create single attosecond electron pulses rather than trains [10]. The simplicity of the system that involves only two electron wavepackets and laser radiation suggests that the proposed system can be used as a test model to better understand exchange and Coulomb correlations and their role in open quantum systems.

## Methods

### Weak electron-electron interactions

We approximate the equation of motion for the second electron as

$$\tilde{v}^H \tilde{\psi}_2 + \tilde{v}^x_{21} \tilde{\psi}_1 = +i\hbar \frac{\partial}{\partial t} \tilde{\psi}_2 \quad , \tag{15}$$

where the exchange potential and Hartree potentials are given by equations (6) and (7). We assume that the dynamics of the first electron is dominated by the interaction with an intense laser field, and therefore, neglect the role of the Coulomb and exchange potentials. We use the Volkov representation to describe the interaction of a free electron with light to obtain $\tilde{\psi}_1 = e^{-i\frac{e}{m_0}\int_0^t \vec{\kappa}_{i1}\cdot\vec{A}(\vec{r},\tau)d\tau}$, where $\vec{A}(\vec{r},t)$ is the scalar potential. Inserting this into (15), we obtain

$$+i\hbar\frac{\partial}{\partial t}\tilde{\psi}_2 = -\frac{e^2}{4\pi\varepsilon_0}e^{i(\vec{\kappa}_{i2}-\vec{\kappa}_{i1})\cdot\vec{r}}e^{-i\frac{e}{m_0}\int_0^t \vec{\kappa}_{i2}\cdot\vec{A}(\vec{r},\tau)d\tau} \times$$

$$\int d^3r' \frac{\tilde{\psi}_2 e^{-i(\vec{\kappa}_{i2}-\vec{\kappa}_{i1})\cdot\vec{r}'+i\frac{e}{m_0}\int_0^t \vec{\kappa}_{i2}\cdot\vec{A}(\vec{r}',\tau)d\tau}}{|\vec{r}-\vec{r}'|}$$

$$+\tilde{\psi}_2 \frac{e^2}{4\pi\varepsilon_0}\int d^3r' \frac{1}{|\vec{r}-\vec{r}'|} \tag{16}$$

that can be recast as

$$i\hbar\nabla^2\left\{\left(\frac{\partial}{\partial t}\tilde{\psi}_2\right)e^{-i(\vec{\kappa}_{i2}-\vec{\kappa}_{i1})\cdot\vec{r}+i\frac{e}{m_0}\int_0^t\vec{\kappa}_{i2}\cdot\vec{A}(\vec{r},\tau)d\tau}\right\}=\frac{e^2}{\varepsilon_0}\tilde{\psi}_2 e^{-i(\vec{\kappa}_{i2}-\vec{\kappa}_{i1})\cdot\vec{r}+i\frac{e}{m_0}\int_0^t\vec{\kappa}_{i2}\cdot\vec{A}(\vec{r},\tau)d\tau}$$
$$+\nabla^2\left\{\tilde{\psi}_2 e^{-i(\vec{\kappa}_{i2}-\vec{\kappa}_{i1})\cdot\vec{r}+i\frac{e}{m_0}\int_0^t\vec{\kappa}_{i2}\cdot\vec{A}(\vec{r},\tau)d\tau}\frac{e^2}{4\pi\varepsilon_0}\int d^3r'\frac{1}{|\vec{r}-\vec{r}'|}\right\} \quad . \tag{17}$$

We can expand the last term and write it as

$$i\hbar\nabla^2\left\{\left(\frac{\partial}{\partial t}\tilde{\psi}_2\right)e^{-i(\vec{\kappa}_{i2}-\vec{\kappa}_{i1})\cdot\vec{r}+i\frac{e}{m_0}\int_0^t\vec{\kappa}_{i2}\cdot\vec{A}(\vec{r},\tau)d\tau}\right\}=\frac{e^2}{\varepsilon_0}\tilde{\psi}_2 e^{-i(\vec{\kappa}_{i2}-\vec{\kappa}_{i1})\cdot\vec{r}+i\frac{e}{m_0}\int_0^t\vec{\kappa}_{i2}\cdot\vec{A}(\vec{r},\tau)d\tau}$$
$$+\nabla^2\left\{\tilde{\psi}_2 e^{-i(\vec{\kappa}_{i2}-\vec{\kappa}_{i1})\cdot\vec{r}+i\frac{e}{m_0}\int_0^t\vec{\kappa}_{i2}\cdot\vec{A}(\vec{r},\tau)d\tau}\right\}\frac{e^2}{4\pi\varepsilon_0}\int d^3r'\frac{1}{|\vec{r}-\vec{r}'|}$$
$$+\nabla^2\left\{\frac{e^2}{4\pi\varepsilon_0}\int d^3r'\frac{1}{|\vec{r}-\vec{r}'|}\right\}\tilde{\psi}_2 e^{-i(\vec{\kappa}_{i2}-\vec{\kappa}_{i1})\cdot\vec{r}+i\frac{e}{m_0}\int_0^t\vec{\kappa}_{i2}\cdot\vec{A}(\vec{r},\tau)d\tau} \tag{18}$$
$$+2\vec{\nabla}\left\{\tilde{\psi}_2 e^{-i(\vec{\kappa}_{i2}-\vec{\kappa}_{i1})\cdot\vec{r}+i\frac{e}{m_0}\int_0^t\vec{\kappa}_{i2}\cdot\vec{A}(\vec{r},\tau)d\tau}\right\}\cdot\frac{e^2}{4\pi\varepsilon_0}\vec{\nabla}\int d^3r'\frac{1}{|\vec{r}-\vec{r}'|}$$

By introducing $\chi=e^{-i(\vec{\kappa}_{i2}-\vec{\kappa}_{i1})\cdot\vec{r}+i\frac{e}{m_0}\int_0^t\vec{\kappa}_{i2}\cdot\vec{A}(\vec{r},\tau)d\tau}$, we obtain:

$$i\hbar\nabla^2\left\{\chi\frac{\partial}{\partial t}\tilde{\psi}_2\right\}=\frac{e^2}{\varepsilon_0}\chi\tilde{\psi}_2$$
$$+\nabla^2\{\chi\tilde{\psi}_2\}\frac{e^2}{4\pi\varepsilon_0}\int d^3r'\frac{1}{|\vec{r}-\vec{r}'|}-\frac{e^2}{\varepsilon_0}\chi\tilde{\psi}_2 \quad , \tag{19}$$
$$+2\vec{\nabla}\{\tilde{\psi}_2\chi\}\cdot\frac{e^2}{4\pi\varepsilon_0}\vec{\nabla}\int d^3r'\frac{1}{|\vec{r}-\vec{r}'|}$$

that can be simplified as

$$i\hbar\nabla^2\left\{\chi\frac{\partial}{\partial t}\tilde{\psi}_2\right\}=\tilde{v}^H\nabla^2\{\chi\tilde{\psi}_2\}+2\vec{\nabla}\{\tilde{\psi}_2\chi\}\cdot\vec{\nabla}\tilde{v}^H \quad . \tag{20}$$

If we assume that $\chi(\vec{r},t)\equiv\chi(t)$, eq. (A6) is reformulated as the well-known Lorentz-force identity as

$$\left(i\hbar\frac{\partial}{\partial t}-\tilde{v}^H\right)\nabla^2\{\tilde{\psi}_2\}=+2\vec{\nabla}\{\tilde{\psi}_2\}\cdot\vec{\nabla}\tilde{v}^H \quad . \tag{21}$$

Equation (21) defines the change in the kinetic energy of the second electron (LHS) due to the electric field caused by the first electron acting on the charged particle (RHS). Obviously, the role of the magnetic field is here neglected since our treatment is nonrelativistic. To better demonstrate the role of time-dependent mutual correlations on the amplitude modulations, we multiply eq. (15) by $\tilde{\psi}_2^*$ and subtract it from its conjugated term to obtain

$$\hbar \frac{\partial}{\partial t} |\tilde{\psi}_2|^2 = 2 \operatorname{Im} \frac{e^2}{4\pi\varepsilon_0} \tilde{\psi}_1^* \tilde{\psi}_2 \int d^3 r' \frac{\tilde{\psi}_1 \tilde{\psi}_2^*}{|\vec{r} - \vec{r}'|} \quad . \tag{22}$$

The phase-differences between the electron wavepackets thus cause time-dependent amplitude modulations.

To obtain an analytical grasp of strong inter-electron interactions (which we simulate fully numerically, see the following section), we take into account the fact that the dynamics of both electrons is affected by the Hartree and exchange potentials, by using $\tilde{v}_2^H \tilde{\psi}_2 + \tilde{v}_{21}^x \tilde{\psi}_1 = +i\hbar \frac{\partial}{\partial t} \tilde{\psi}_2$ and $\tilde{v}_1^H \tilde{\psi}_1^* + \left(\tilde{v}_{12}^x\right)^* \tilde{\psi}_2^* = -i\hbar \frac{\partial}{\partial t} \tilde{\psi}_1^*$. We multiply the former by $\tilde{\psi}_1^*$ and the latter by $\tilde{\psi}_2$ and subtract them to obtain

$$i\hbar \frac{\partial}{\partial t} C(\vec{r}, t) = \frac{e^2}{4\pi\varepsilon_0} \left\{ C(\vec{r}, t) \int d^3 r' \frac{\Delta(\vec{r}', t)}{|\vec{r} - \vec{r}'|} - \Delta(\vec{r}, t) \int d^3 r' \frac{C(\vec{r}', t)}{|\vec{r} - \vec{r}'|} \right\} \tag{23}$$

where $\Delta(\vec{r}, t) = |\psi_1|^2 - |\psi_2|^2$, and $C(\vec{r}, t) = \tilde{\psi}_1^*(\vec{r}, t) \tilde{\psi}_2(\vec{r}, t)$. We can thus derive a similar equation for the time-dependent amplitude modulations, given by

$$\hbar \frac{\partial}{\partial t} \Delta(\vec{r}, t) = 4 \operatorname{Im} \frac{e^2}{4\pi\varepsilon_0} C(\vec{r}, t) \int d^3 r' \frac{C^*(\vec{r}', t)}{|\vec{r} - \vec{r}'|} . \tag{24}$$

Therefore, there is a strong relation between the mutual phase correlations and differential amplitude modulations.

**Maxwell-Schrödinger simulations**

For numerically calculating the dynamics of the electrons, equations (5) to (7) are used and combined with a Maxwell solver that is based on the finite-difference time-dependent method [33]. The simulations are performed in a two dimensional (2D) cartesian coordinate system, however, a screened potentials is used to correctly model the three-dimensional (3D) Coulomb potential. In order to calculate the Hartree and exchange potentials in 2D space, we have used the Poisson equation in the form of $\varepsilon_0 \nabla^2 \tilde{v}_m^H = -e^2 |\psi_m(\vec{r}', t)|^2$ and $\varepsilon_0 \nabla^2 \left( \tilde{v}_{nm}^x e^{-i(\vec{\kappa}_m - \vec{\kappa}_n)\cdot\vec{r}} \right) = e^2 \tilde{\psi}_m^*(\vec{r}, t) \tilde{\psi}_n(\vec{r}, t) e^{-i(\vec{\kappa}_m - \vec{\kappa}_n)\cdot\vec{r}}$, for the Coulomb and exchange potentials, respectively, and adopted the Fourier method for spatial differentiation [34]. The transformation from the 2D to the 3D potential is done by considering a confinement of 3.3 nm along the *yz*-transverse plane. We have compared our numerically calculated Hartree and exchange potentials with analytical solutions including Gaussian charge distributions and have found a very good agreement. The full details on the developed numerical toolbox can be found in [12, 21, 26]. The TDHF simulations are written in the MATLAB environment and are operated on a supercomputing node with 17 CPUs and 125 GB RAM. The simulation time is approximately 36 hours. A 3D version of this code takes 14 days, and we did not observe a significant deviation from the results shown in Fig. 2, thanks to the symmetry of the structure and the excitations. Therefore, we continued with our 2D analysis.

**Exchange in two-particle density matrices**

For a two-particle system described by the wavefunction $|\psi\rangle$ the two-particle density (pair-probability density) is given by

$$\rho_{12}(\vec{r}_1, \vec{r}_2) = 2 \sum_{s_1, s_2} \langle \vec{r}_1 s_1, \vec{r}_2 s_2 | \psi \rangle \langle \psi | \vec{r}_1 s_1, \vec{r}_2 s_2 \rangle, \tag{25}$$

where $s_i$ denotes the spin degrees of freedom. The factor 2 comes from the here used normalization of the pair-probability density to $N(N-1)$ which is 2 in our case. Tracing out one further particle leads to the particle-density given by

$$\rho_1(\vec{r}_1) = \int d^3 r_2 \rho_{12}(\vec{r}_1, \vec{r}_2). \tag{26}$$

(Again the particle density is normalized here to the particle number which is 2.)

This leads to

$$\rho_1(\vec{r}_1) = |\psi_1(\vec{r}_1)|^2 + |\psi_2(\vec{r}_1)|^2, \tag{27}$$

for both cases (spin polarized and un-polarized). The two-particle density $\rho_{12}(\vec{r}_1, \vec{r}_2)$ can now be decomposed in the following way:

$$\rho_{12}(\vec{r}_1, \vec{r}_2) = \rho_1(\vec{r}_1)\rho_1(\vec{r}_2) + \rho_x(\vec{r}_1, \vec{r}_2) + \rho_c(\vec{r}_1, \vec{r}_2). \tag{28}$$

The first part is the completely uncorrelated contribution, the second part is the contribution coming from the exchange term, and the last term originates from the two-particle cumulant and measures particle-correlations beyond a single Slater determinant (not treated in the present paper). By construction this contribution is 0 if one makes just a Hartree-Fock ansatz for the two-particle wavefunction. However, in general, it is non-negligible. The term $\rho_x(\vec{r}_1, \vec{r}_2)$ is usually used to measure exchange correlations. For our two-electron system it is given by

$$\rho_x(\vec{r}_1, \vec{r}_2) = -|\psi_1(\vec{r}_1)|^2 |\psi_1(\vec{r}_2)|^2 - |\psi_2(\vec{r}_1)|^2 |\psi_2(\vec{r}_2)|^2$$
$$-\psi_1^*(\vec{r}_1)\psi_2^*(\vec{r}_2)\psi_2(\vec{r}_1)\psi_1(\vec{r}_2) - \psi_2^*(\vec{r}_1)\psi_1^*(\vec{r}_2)\psi_1(\vec{r}_1)\psi_2(\vec{r}_2), \tag{29}$$

for the polarized case, and by

$$\rho_x(\vec{r}_1, \vec{r}_2) = -|\psi_1(\vec{r}_1)|^2 |\psi_1(\vec{r}_2)|^2 - |\psi_2(\vec{r}_1)|^2 |\psi_2(\vec{r}_2)|^2, \tag{30}$$

for the unpolarized case. Since in both cases the electrons are entangled, anti-correlations are present independent of polarization. However, in the unpolarized case they are mediated only be products of orbital densities and lead to an overall phase-independent reduction of the uncorrelated two-particle density. In the main text, we have, therefore, focused on the phase-dependent part that distinguishes $\rho_x$ in the polarized case from $\rho_x$ in the unpolarized case. Finally, we briefly mention that we have carefully compared our results with those resulting from orthogonal wavepackets stemming from $|1'\rangle = |1\rangle$ and

$$|2'\rangle = \left(|2\rangle - \langle 1|2\rangle |1\rangle\right)\left(1 - |\langle 1|2\rangle|^2\right)^{-1}$$

and they match within an accuracy of 0.001%.

**Acknowledgement**

The authors gratefully thank Michael Bonitz (Kiel University) for fruitful discussions. This project has received funding from the European Research Council (ERC) under the European Union's Horizon 2020 research and innovation program, Grant Agreements No. 802130 (Kiel, NanoBeam) and Grant Agreements No. 101017720 (EBEAM). Financial support from Deutsche Forschungsgemeinschaft under the Art. 91 b GG Grant Agreement No. 447330010 and Grant Agreement No. 440395346 is acknowledged.